\documentclass[conference]{IEEEtran}
\IEEEoverridecommandlockouts
\usepackage{cite}
\usepackage{amsmath,amssymb,amsfonts}
\usepackage{algorithmic}
\usepackage{graphicx}
\usepackage{textcomp}
\usepackage{xcolor}
\usepackage[numbers,sort&compress]{natbib}
\usepackage{float}
\usepackage{inputenc}
\usepackage{amsmath}
\usepackage[ruled]{algorithm2e}
\def\BibTeX{{\rm B\kern-.05em{\sc i\kern-.025em b}\kern-.08em
    T\kern-.1667em\lower.7ex\hbox{E}\kern-.125emX}}
\begin{document}

\title{A Novel Approach to Malicious Code Detection Using CNN-BiLSTM and Feature Fusion\\

\thanks{Identify applicable funding agency here. If none, delete this.}
}

\author{\IEEEauthorblockN{1\textsuperscript{st} Lixia Zhang}
\IEEEauthorblockA{\textit{XI'AN XD POWER SYSTEMS CO.,LTD.}\\
Xi'an,China \\
23151214343@stu.xidian.edu.cn}
\and
\IEEEauthorblockN{2\textsuperscript{nd} Tianxu Liu*}
\IEEEauthorblockA{\textit{Hangzhou Research Institute of Xidian University}\\
Hangzhou,China \\
23151214343@stu.xidian.edu.cn}
\and
\IEEEauthorblockN{3\textsuperscript{rd} Kaihui Shen}
\IEEEauthorblockA{\textit{Xidian digital technology}\\
Xi'an,China \\
23151214343@stu.xidian.edu.cn}
\and
\IEEEauthorblockN{4\textsuperscript{th} Cheng Chen}
\IEEEauthorblockA{\textit{ Harbin Engineerning University}\\
Harbin,China\\
23151214343@stu.xidian.edu.cn}
}

\maketitle

\begin{abstract}
With the rapid advancement of Internet technology, the threat of malware to computer systems and network security has intensified. Malware affects individual privacy and security and poses risks to critical infrastructures of enterprises and nations. The increasing quantity and complexity of malware, along with its concealment and diversity, challenge traditional detection techniques. Static detection methods struggle against variants and packed malware, while dynamic methods face high costs and risks that limit their application. Consequently, there is an urgent need for novel and efficient malware detection techniques to improve accuracy and robustness.

This study first employs the minhash algorithm to convert binary files of malware into grayscale images, followed by the extraction of global and local texture features using GIST and LBP algorithms. Additionally, the study utilizes IDA Pro to decompile and extract opcode sequences, applying N-gram and tf-idf algorithms for feature vectorization. The fusion of these features enables the model to comprehensively capture the behavioral characteristics of malware.

In terms of model construction, a CNN-BiLSTM fusion model is designed to simultaneously process image features and opcode sequences, enhancing classification performance. Experimental validation on multiple public datasets demonstrates that the proposed method significantly outperforms traditional detection techniques in terms of accuracy, recall, and F1 score, particularly in detecting variants and obfuscated malware with greater stability.

The research presented in this paper offers new insights into the development of malware detection technologies, validating the effectiveness of feature and model fusion, and holds promising application prospects.
\end{abstract}


\section{Introduction}
With the rapid development of Internet technology, malware poses an increasingly severe threat to global cybersecurity \cite{kumar2023emerging}. Malware not only compromises individual users' privacy but also endangers business continuity, secure transactions in financial institutions, and the operation of critical national infrastructures \cite{lis2019cyberattacks}. The diversity, complexity, and stealth of malware make its detection and defense one of the most challenging tasks in the field of cybersecurity. In recent years, the number of malware instances has grown exponentially, with attack methods constantly evolving. Particularly, malware variants and polymorphism techniques have significantly increased detection difficulty \cite{thakur2024cyber}.

Currently, malware detection techniques are broadly divided into two categories: static analysis and dynamic analysis. Static analysis involves parsing the binary files, instruction sequences, or assembly code of malware, relying on feature extraction and matching for classification \cite{wolsey2022stateoftheartaibasedmalwaredetection}. Its advantages include high detection speed and the ability to analyze code without execution. However, static analysis struggles against code obfuscation, packing techniques, and malware variants, making it susceptible to evasion \cite{app12178482}. Dynamic analysis, on the other hand, executes malware in a virtual environment, recording its behavior for analysis \cite{li2020advanced}. Although dynamic analysis effectively handles obfuscated code and identifies malware behavior, its high computational cost and time requirements limit its scalability \cite{Ngo2022Fast}. Additionally, dynamic analysis faces the challenge of avoiding detection by malware, which may prevent the method from functioning effectively in certain cases \cite{Aboaoja2023Dynamic}.

In response to these challenges, machine learning and deep learning techniques have recently been applied to malware detection, aiming to enhance detection accuracy and robustness through automated feature extraction and model training \cite{Brown2023Automated}. Compared to traditional methods, deep learning can handle more complex feature spaces and is well-suited for modeling large-scale, multi-dimensional data \cite{Shukla2019Stealthy}. However, existing deep learning-based malware detection methods still suffer from limitations in feature extraction, as most rely on either static or dynamic features, making it difficult to fully capture the diversity of malware behavior \cite{Al-Hashmi2022Deep-Ensemble}. As a result, combining multiple feature sources and effectively integrating them has become a key focus in current research \cite{Gibert2022Fusing}.

To address these challenges, we present a deep learning-based static malware detection method that fuses image texture features with opcode sequence features using a CNN and a Bidirectional Long Short-Term Memory (BiLSTM) architecture. The minhash algorithm converts malware binary files into grayscale images, reducing data complexity for effective feature extraction. GIST extracts global texture features, while Local Binary Pattern (LBP) captures local texture features, enhancing the characterization of similarities and differences among malware types.

The opcode sequence of malware is extracted using the IDA Pro disassembler and processed with N-gram and tf-idf algorithms, allowing the model to capture behavioral patterns and execution logic alongside image features. This creates a multimodal input dataset for comprehensive malware recognition.
We propose a CNN-BiLSTM fusion model to enhance detection. The CNN extracts local and global features from malware images, while the BiLSTM captures contextual dependencies from opcode sequences. Jointly modeling these features significantly improves the identification of complex malware behaviors, resulting in more accurate detection.

In the experimental section, we conduct comprehensive evaluations on several public malware datasets using metrics such as accuracy, recall, and F1 score. The results show that the proposed fusion-based detection method significantly outperforms traditional techniques, especially in detecting malware variants and obfuscated samples, demonstrating higher stability and robustness.

The contributions of this paper are primarily reflected in the following aspects:

Feature-level contribution: This paper proposes a method that fuses image texture features with opcode features, overcoming the limitations of single-feature extraction and achieving multidimensional capture of static malware characteristics.

Model-level contribution: A fusion model combining CNN and BiLSTM is designed and implemented, leveraging the strengths of both in handling image and sequence data, which significantly improves the accuracy and efficiency of malware detection.

Experiment-level contribution: Through experiments conducted on multiple datasets, the proposed model demonstrates excellent performance in malware detection tasks, offering new insights and methods for future malware detection technologies.

The structure of this paper is as follows: Section 2 reviews related work on existing malware detection techniques; Section 3 details the proposed feature extraction and model design methods; Section 4 presents the experimental setup, results, and analysis; and finally, Section 5 concludes the research and discusses future work.

\section{Related Work}

Malware detection techniques have evolved from traditional signature-based methods to sophisticated machine learning and deep learning approaches. This section reviews key developments in static analysis, dynamic analysis, deep learning-based detection, and hybrid methods that enhance detection accuracy.

\subsection{Traditional Malware Detection Techniques}
Static analysis is a common malware detection method that inspects binary code without execution. Traditional techniques, like signature-based detection, compare known malware signatures or hash values against a database \cite{10.1007/978-3-030-37309-2_3}. While effective against known threats, this method struggles to detect unknown or modified malware, especially those using code obfuscation, packing, or encryption to disguise their behavior \cite{info15020102}.To address these limitations, researchers have proposed advanced static analysis techniques, including control flow, opcode sequence, and API call analysis, to capture higher-level structural and semantic features of malware.

Opcode sequence analysis focuses on specific instruction sequences in the binary code, enabling the detection of behaviorally similar malware variants with shared operational characteristics.Despite these improvements, static analysis still struggles to detect polymorphic and metamorphic malware, which change their structure dynamically to evade detection \cite{faruki2023survey}.

Dynamic analysis executes malware in a controlled environment (e.g., a sandbox) to monitor its real-time behavior. By observing system calls, file manipulations, and network communications, it detects previously unseen malware based on its actions rather than static features.This makes dynamic analysis particularly effective against obfuscated malware. Tools like Cuckoo Sandbox have become popular for automating dynamic malware analysis \cite{ilic2022pilot}.

Dynamic analysis faces challenges such as high computational costs and time demands for executing malware in a virtual environment. Modern malware often employs anti-analysis techniques like virtual machine detection or execution delays, effectively evading detection. Thus, dynamic analysis alone is often inadequate for large-scale malware detection \cite{maniriho2022study}.

\subsection{Machine Learning and Deep Learning in Malware Detection}
In recent years, machine learning (ML) and deep learning (DL) have become powerful tools for malware detection, enabling automatic extraction of complex patterns from data without the need for manually engineered features. Traditional machine learning models, such as Support Vector Machines (SVMs), Decision Trees, and Random Forests, are widely used in malware detection tasks \cite{singh2021survey}. These models rely on manually selected features, such as opcode frequencies, byte sequences, and system call traces, to train classifiers that distinguish between benign and malicious files \cite{gopinath2023comprehensive}.

While these classical ML methods have achieved success, they require significant domain expertise to define effective features, and their performance is often limited by the quality and completeness of the feature set \cite{liu2024transurl}. In contrast, deep learning models like Convolutional Neural Networks (CNNs) and Recurrent Neural Networks (RNNs) can automatically learn high-level features from raw data, enhancing their adaptability and scalability to large datasets \cite{shiri2023comprehensiveoverviewcomparativeanalysis}.

CNNs have been successfully applied to malware detection by converting malware binaries into grayscale images, with pixel values representing byte patterns in the binary file \cite{Omar2022}. This allows CNNs to capture spatial patterns in malware, indicative of malicious behavior, similar to their use in image recognition tasks \cite{gibert2019using}. Studies show that CNNs can achieve high accuracy in distinguishing between benign and malicious software by analyzing image-based representations of malware \cite{app13074624}.

RNNs, particularly Long Short-Term Memory (LSTM) networks and Bidirectional LSTMs (BiLSTMs), are applied to sequential data like opcode sequences, system call traces, and network traffic logs \cite{pu2024larod}. These models capture temporal dependencies, modeling malware's dynamic behavior over time. LSTM-based models, for example, detect malicious behaviors across multiple execution steps, offering a comprehensive view of malware operations \cite{qian2024mufuzz}.

Despite deep learning's success in feature extraction and detection, most methods focus on static or dynamic features alone, limiting their ability to capture malware's complexity. This has driven interest in feature fusion techniques to improve accuracy and robustness \cite{johny2024deep,jia2024ethereum}.

\subsection{Hybrid Detection Methods and Feature Fusion}

To address single feature limitations, recent research has explored hybrid detection methods, combining static and dynamic features for a more comprehensive view of malware \cite{math11132944}.Some studies combine static features, like opcode sequences and byte n-grams, with dynamic features, such as system calls and network logs, improving detection accuracy and resilience against obfuscated malware \cite{10.1007/978-3-030-91356-4_14}.

A related line of research focuses on multi-modal feature fusion, which seeks to combine diverse data sources into a unified model \cite{zhu2021malware}. Image-based features from malware binaries can be fused with opcode sequence features to provide a comprehensive view of malware's structural and behavioral properties \cite{hashemi2023ifmd}. This approach effectively handles malware variants by reducing reliance on a single feature type, making the model more robust to evasion techniques.

A key example is CNN-BiLSTM models, which combine CNNs' strength in extracting spatial features from malware images with BiLSTMs' ability to model opcode sequences \cite{akhtar2022detection}. These hybrid models excel in detecting malware, especially packed or metamorphic variants where traditional methods struggle. By fusing features from multiple modalities, they significantly enhance the robustness and generalization of malware detection systems.
\section{Methodology}
This section details the proposed deep learning-based static malware detection method, organized into three key components: feature extraction, feature fusion, and model construction. As shown in Figure 1, the feature extraction stage focuses on obtaining static features from malware, including grayscale texture features from binary files and opcode features through reverse engineering. These features offer comprehensive insights into the malware's global structure (image features) and local behavior (opcode sequences).
\begin{figure}[H]
    \centering
    \includegraphics[width=0.7\linewidth]{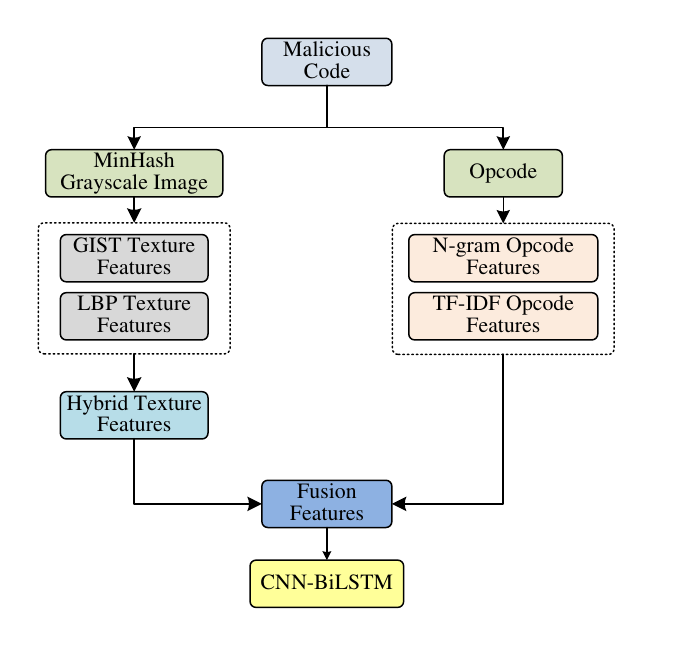}
    \caption{Overall flow chart of the program}
    \label{fig:enter-label}
\end{figure}
In the feature fusion stage, extracted texture and opcode features are combined to enhance feature representation. This integration enables the model to capture broader malware characteristics, addressing the limitations of traditional methods that rely on a single feature source.

The model construction stage develops a CNN-BiLSTM fusion model, combining convolutional neural networks (CNN) for global image pattern recognition and bidirectional long short-term memory networks (BiLSTM) for sequential opcode data dependencies. This integration effectively processes both global and local information, improving detection accuracy and robustness. Comparative experiments demonstrate the proposed method's effectiveness in detecting various malware types, including polymorphic and obfuscated samples.
\subsection{Feature Extraction}
Feature extraction is key for static malware detection and greatly impacts model performance. This paper combines grayscale image texture features with opcode sequence features to capture both global and local characteristics. The process includes data acquisition, preprocessing, grayscale image generation, and texture and opcode feature extraction.
\subsubsection{Data Acquisition and Preprocessing}
The quality of the dataset directly impacts feature extraction effectiveness. In this paper, we utilize the malware dataset from the VirusShare platform, which includes samples uploaded from September 2023 to February 2024, primarily consisting of executable files (EXE) for the Windows platform. 
Since the original dataset contains various file types, such as PDFs and DLLs, we first filtered it to retain only Portable Executable (PE) format files. The data filtering process involved using the VirusShare API to obtain the MD5 hash list of the malware samples, followed by recognizing each file's format based on its MD5 hash and selecting only PE format files. Finally, we deduplicated the selected samples to ensure dataset uniqueness and accuracy. After filtering and labeling, we obtained a dataset containing 14 malware families with a total of 12,021 samples, as shown in Table 1, which lists the sample distribution of each family.
\begin{table}[H]
\caption{Distribution of sample size by family}
\begin{center}
\label{tab:1}
\begin{tabular}{|c|c|c|}
\hline
\textbf{Index} & \textbf{Malware Family Name} & \textbf{Number of Samples} \\ \hline
1              & Ramnit                       & 1336                       \\ \hline
2              & Spyware                      & 890                        \\ \hline
3              & Kelihos\_ver3                & 1923                       \\ \hline
4              & Vundo                        & 469                        \\ \hline
5              & Banload                      & 211                        \\ \hline
6              & Kelihos\_ver1                & 379                        \\ \hline
7              & Obfuscator.ACY               & 1307                       \\ \hline
8              & Gatak                        & 993                        \\ \hline
9              & Bundlore                     & 158                        \\ \hline
10             & Zeroaccess                   & 192                        \\ \hline
11             & Qhost                        & 132                        \\ \hline
12             & Downloader                   & 1261                       \\ \hline
13             & Tracur                       & 781                        \\ \hline
14             & Lollipop                     & 1989                       \\ \hline
\end{tabular}
\end{center}
\end{table}
\subsubsection{Grayscale Image Generation}
A grayscale image is represented using only the grayscale channel, consisting of three colors: black, white, and shades of gray (ranging from 0 to 255). Using grayscale images effectively reduces data complexity, shortens processing time, and mitigates distortion issues associated with more complex images. After converting malicious code into grayscale images, as shown in Figure 2, malware from the same family exhibits similar texture features, while malware from different families shows distinct differences.
\begin{figure}[H]
    \centering
    \vspace{-0.5cm}
    \includegraphics[width=0.5\linewidth]{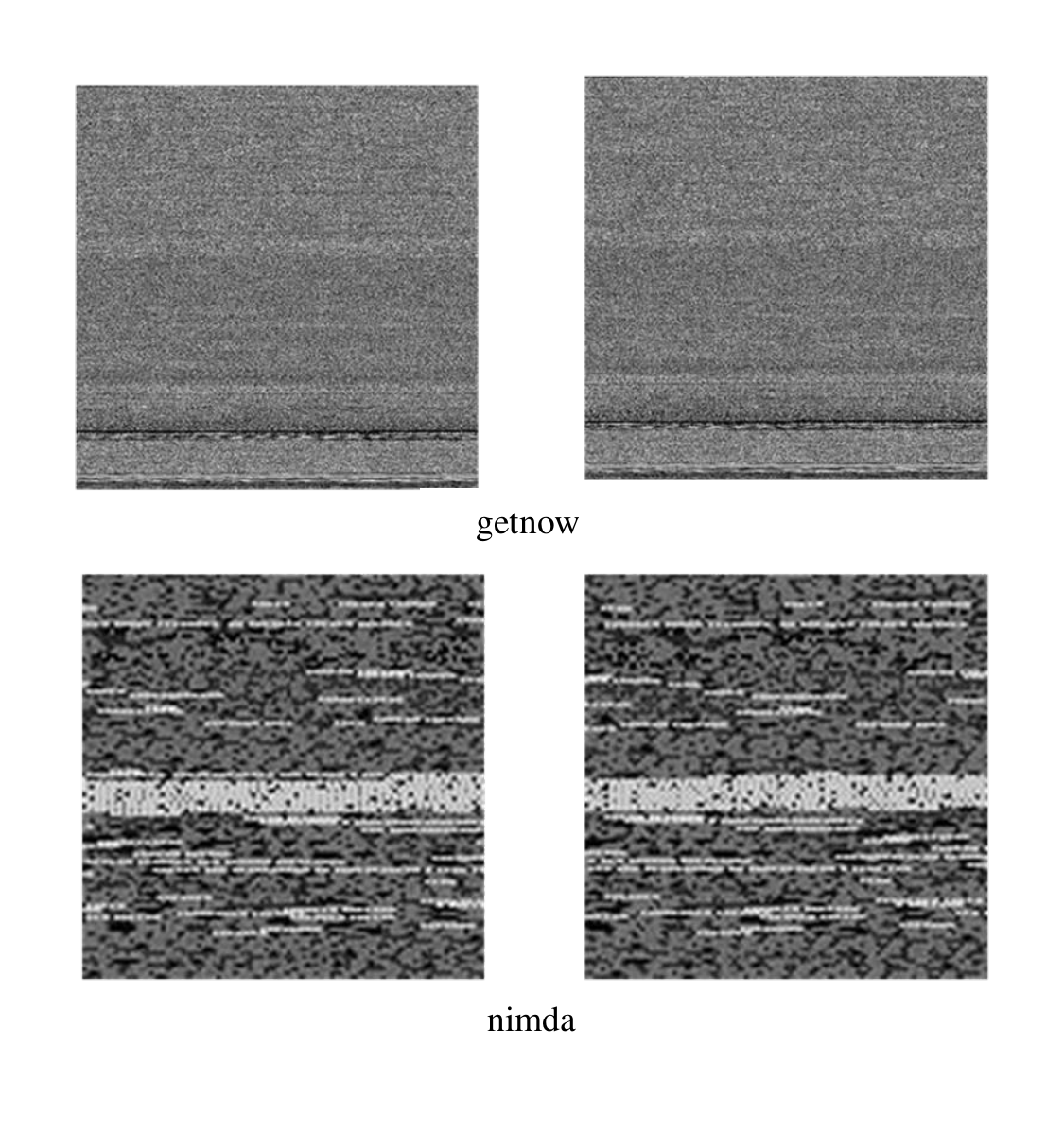}
    \caption{The getnow family versus the nimda family}
    \label{fig:enter-label}
\end{figure}
During the conversion to a grayscale image, the binary file of the malicious code is processed byte by byte, with each 8-bit segment mapped directly to a pixel. This method creates a two-dimensional grayscale image where each byte value (0-255) corresponds to a pixel's grayscale value.

Before generating the grayscale image, the opcode sequence is extracted from the executable file using the decompilation tool IDA Pro, which produces assembly (asm) files. Opcodes are extracted through batch processing with Python scripts, identifying the ".text" segment to locate and extract the opcode sequences. The Minhash algorithm is then used to generate hash signatures.

To convert these hash signatures into grayscale images, as shown in Figure 3, hash values (typically exceeding 256x256) are processed by dividing by 256 and applying the modulus operation. This maps the hash values to x and y coordinates on a two-dimensional plane, with the grayscale value (z) corresponding to the modulus result. The image size is set to 128x128, balancing information retention and computational efficiency. Each malicious code hash signature is thus transformed into a grayscale image, with each pixel representing a grayscale value derived from the hash signature.
\begin{figure}[H]
    \centering
    \includegraphics[width=0.6\linewidth]{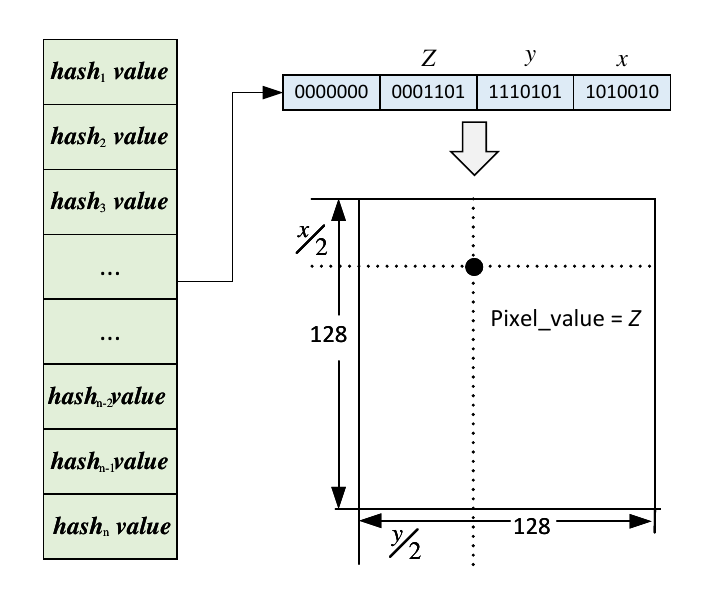}
    \caption{Schematic diagram of grayscale image conversion process}
    \label{fig:enter-label}
\end{figure}
We employs the GIST and LBP algorithms to extract global and local texture features from the grayscale images. The GIST algorithm captures global features by applying Gabor filters and dividing the image into sub-blocks. Each sub-block undergoes convolution to generate GIST features, reflecting the overall spatial layout and texture characteristics of the malicious code, as illustrated in Figure 4.
\begin{figure}[H]
    \centering
    \includegraphics[width=1\linewidth]{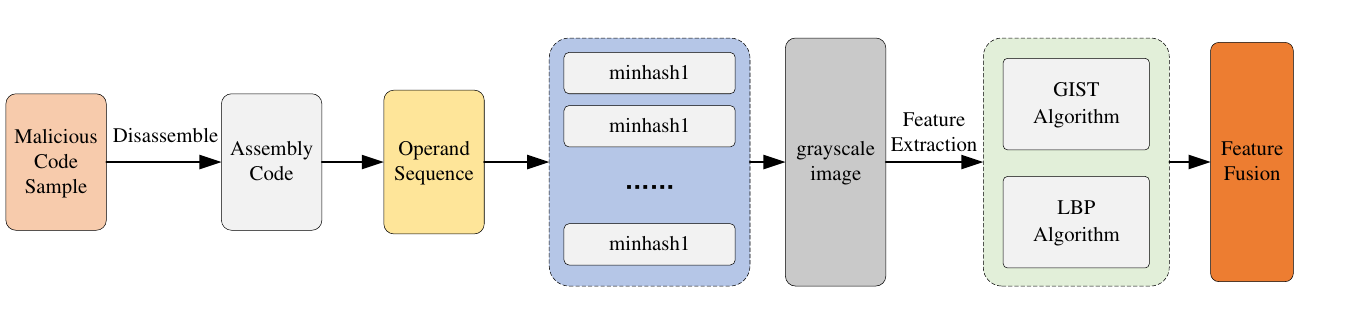}
    \caption{Grayscale Image Texture Feature Extraction Process}
    \label{fig:enter-label}
\end{figure}

\subsubsection{Texture Feature Extraction}
The GIST algorithm is used to extract global texture features from an image by applying convolution with multi-scale and multi-directional Gabor filters to capture its spatial layout and global patterns. The grayscale image is divided into regular sub-blocks, with each sub-block representing a local area. Convolution operations are performed on these sub-blocks using various Gabor filters of different scales and orientations to extract global features. The feature vectors from each sub-block are then concatenated to form the GIST features for the entire grayscale image, capturing the global patterns of the malicious code.

Gabor filters effectively extract edge, texture, and spatial frequency information by operating in both spatial and frequency domains, reflecting the image's overall layout. Specifically, the GIST algorithm's workflow involves filtering the grayscale image with Gabor filters, dividing the filtered image into sub-blocks, and calculating the mean or other statistical metrics for each sub-block to obtain its feature vector. The mathematical representation is as follows:
\begin{equation}
    G_{i j}=\sum_{k=1}^{K} f_{k}\left(I_{i j}\right)\label{eq}
\end{equation}

In this context, $G_{j}^{i}$ represents the GIST feature of the image at position \((i,j)\) , $f_{k}$ denotes the k-th Gabor filter, and $I_{j}^{i}$ is the pixel value of the image. Through this approach, the GIST features can capture the global spatial layout information of the grayscale image, thereby reflecting the global characteristics of the malicious code.

Next, the LBP (Local Binary Patterns) algorithm is used to extract local texture features from an image. LBP is a local texture descriptor that compares grayscale values of each pixel with its neighboring pixels. 
The process involves selecting P neighboring pixels around each pixel and comparing their grayscale values to that of the center pixel. If a neighboring pixel's value is greater than or equal to the center pixel's value, it is assigned a value of 1; otherwise, it is assigned 0. These binary values combine to form a binary number representing the LBP value for that pixel. After traversing the entire image, the LBP values create the local feature description.
LBP is rotationally and grayscale invariant, making it highly effective for describing local texture features, particularly for detecting edges and corners. The computational procedure of the LBP algorithm can be summarized as follows:
\begin{equation}
    L B P\left(x_{c}, y_{c}\right)=\sum_{p=0}^{P-1} s\left(g_{p}-g_{c}\right) \cdot 2^{p}\label{eq}
\end{equation} 

Among other things, the $g_{p}$ is the gray value of the neighborhood pixel. $g_{c}$ is the gray value of the center pixel and the function s(x) is defined as:
\begin{equation}
    s(x)=\left\{\begin{matrix}  1& ifx>0\\  0& otherwise\end{matrix}\right.\label{eq}
\end{equation}

LBP extracts local texture features by analyzing the relationship between each pixel and its surrounding pixels, making it effective for capturing edges and local details in an image. The local features derived from LBP reflect variations in malicious code at a detailed level, aiding in the differentiation of various strains and families of malware.

After extracting GIST and LBP features, the paper integrates these two types to form the final texture representation. GIST provides global spatial layout information, while LBP captures local texture details. This combination enables a comprehensive characterization of malicious code from multiple perspectives. By integrating global and local features, the distinctiveness of the texture characteristics is enhanced, offering rich feature data for subsequent classification and detection of malicious code.
\subsubsection{Opcode Sequence Extraction}
To accurately represent malware's behavioral characteristics through static analysis, this paper employs the IDA Pro disassembly tool to convert binary files into assembly code sequences. By analyzing the Control Flow Graph (CFG), we obtain function call relationships, which facilitate the extraction of opcode sequences.
The opcode extraction process begins at the root node of the CFG, recursively traversing each sub-node to collect opcodes in the order of function execution. For example, as shown in Figure 5, traversing the CFG yields the opcode sequence: {push, mov, call, xor, push, push, ...}. These opcodes represent the program's instructions, reflecting its behavioral logic.
\begin{figure}[H]
    \centering
    \includegraphics[width=0.6\linewidth]{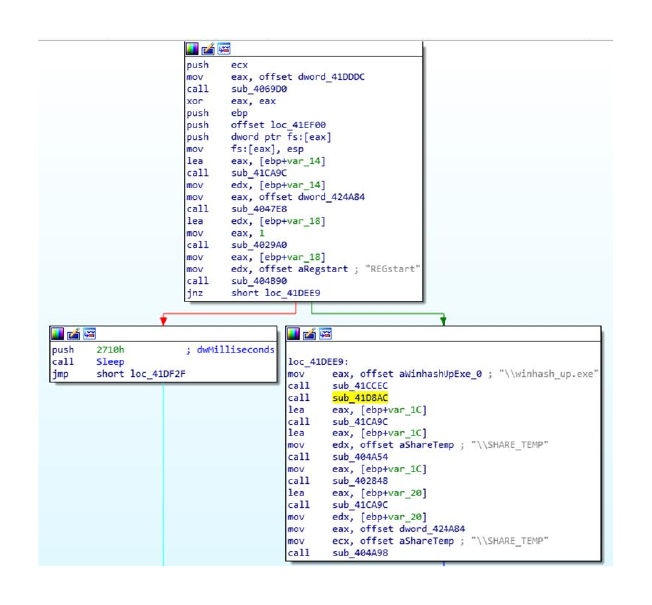}
    \caption{Function Call Relationships}
    \label{fig:enter-label}
\end{figure}
Directly utilizing the extracted opcode sequences for malware classification presents certain limitations, as opcodes are high-dimensional and unstructured data that cannot be directly input into machine learning models. Thus, this paper further processes the opcode sequences for feature representation, primarily employing N-gram and tf-idf algorithms. The combination of these two algorithms enables the capture of both local sequential patterns and the global significance of opcodes from different dimensions, thereby providing effective input features for subsequent modeling.

The N-gram algorithm is a commonly used method for sequence feature extraction, capable of preserving the contextual relationships within the opcode sequence by dividing it into smaller subsequences to extract local features. In this paper, the value of N is set to 3, i.e., a 3-gram model is used to process the opcode sequence. Assuming the opcode sequence is \{call, call, push, call, add, mov, xor\}, the N-gram algorithm generates the following subsequence set: \{(call, call, push), (call, push, call), (push, call, add), (call, add, mov), (add, mov, xor)\}. These subsequences capture the local dependencies between adjacent opcodes, revealing short-range patterns in the execution logic of the malware.

To convert these N-gram subsequences into vectors that can be processed by machine learning models, each subsequence is binarized: if the subsequence exists in a specific malware sample, the corresponding position in the feature vector is marked as 1; otherwise, it is marked as 0. The feature vectors generated by this method not only express the contextual dependency information within the opcode sequence but also ensure the sparsity and processability of the features. The detailed steps of the N-gram algorithm are shown in Algorithm 1.

\begin{algorithm}
\caption{N-gram Opcode Feature Extraction Algorithm}
\KwIn{$n$ (length), $opcode\_seq$ (opcode sequence)}
\KwOut{$q$ (feature vector)}

1. Define and initialize the subsequence set $S$ \;
2. Define variable $m$, the number of subsequences, $m = \text{len}(opcode\_seq) - n + 1$ \;
3. Define feature vector $q$, initialize it as an all-zero vector of length $m$ \;
4. $S = n\text{-gram}(opcode\_seq)$ to get the subsequence set \;
5. \For{$v_i \in S$}{
6. \If{$v_i \in S$}{
    $q[i] = 1$\;
}
\Else{
    $q[i] = 0$\;
}
}
\end{algorithm}
To further enhance the representation of opcode features, this paper introduces the tf-idf algorithm to weight the significance of each opcode. The Term Frequency-Inverse Document Frequency (tf-idf) algorithm is capable of evaluating the relative importance of an opcode in a given malware sample. Specifically, the tf-idf value is proportional to the frequency of the opcode in the sample, and inversely proportional to its frequency across the entire dataset. If an opcode appears frequently in a specific malicious program but rarely in others, it holds higher discriminative power for identifying that malware. In this work, the tf-idf value is computed for each opcode in the sequence, and a weight vector is generated for the model to utilize. The algorithmic procedure for tf-idf feature extraction is shown in Algorithm 2.

\begin{algorithm}
\caption{tf-idf Opcode Feature Extraction Algorithm}
\KwIn{$opcode\_seq$ (opcode sequence), $documents$ (document corpus)}
\KwOut{$v$ (tf-idf feature vector)}

1. Define a variable $v$, of list type, to store tf-idf feature values, initialized as an empty list \;
2. \For{$opcode \in opcode\_seq$}{
    3. $tf = \text{opcode\_seq.count}(opcode)$ to get the term frequency of the current opcode in the sequence \;
    4. Define $count = 0$, to record the number of documents containing the opcode \;
    5. \For{$document \in documents$}{
        6. \If{$opcode \in document$}{
            7. $count += 1$ \;
        }
    }
    8. $idf = \log \left( \frac{\text{len}(documents) + 1}{count + 1} \right)$ \;
    9. $score = tf * idf$ \;
    10. $v.\text{append}(score)$ \;
}
\end{algorithm}
By combining the N-gram and tf-idf algorithms, this paper is able to extract the features of the opcodes from both local and global levels, ensuring that the sequence features and weight information of the opcodes can be effectively fused, and providing high-quality feature inputs for the subsequent training of the model.
\subsection{Feature Fusion}
In the feature fusion stage, we combine extracted image texture features with opcode features. First, we load the feature vectors from the GIST and LBP algorithms, along with those from the N-gram and tf-idf methods. We perform dimensionality checks to ensure each feature set contains the same number of samples. Using the pandas library in Python, we concatenate the texture and opcode feature vectors with the merge() function. Specifically, we merge GIST and LBP features to create a combined texture representation, followed by merging N-gram features with tf-idf features for a comprehensive opcode representation. 

This process creates a fused feature vector that captures both global and local characteristics, along with contextual opcode information. We then merge this vector with its corresponding labels to form a complete dataset. To improve training effectiveness, we standardize the fused feature vector using z-score normalization, which reduces the impact of varying feature dimensions and aids model convergence.

The final labeled fused feature vector is used as input for the deep learning model during training and evaluation. This multi-feature fusion approach enhances the accuracy and robustness of static malware detection. Experimental results demonstrate that the classification accuracy of the fused features significantly surpasses that of single-feature inputs, validating the effectiveness of the fusion strategy.

\subsection{Model Design}
In the preceding section, fused features were obtained by integrating two types of features for input. This subsection employs the integrated deep learning model CNN-BiLSTM for static malware identification. We combine Convolutional Neural Networks (CNNs) and Bidirectional Long Short-Term Memory (BiLSTM) networks. Texture features are extracted from grayscale representations of the malware, leveraging CNNs' ability to process image features through convolutional kernels that extract local features while preserving spatial information and exhibiting translational invariance. Given the inherent sequential dependencies in opcode sequences and the fused features, we utilize the BiLSTM model, which captures both forward and backward contexts to effectively extract deep-level long-range dependencies and latent relational features in malware.

This section adopts the CNN-BiLSTM fusion model, using CNN to enhance the spatial understanding of local features extracted from images and BiLSTM to capture contextual information within opcode sequences. Finally, the fused features are connected through a fully connected layer and classified using the softmax activation function for output.

\subsubsection{CNN Layer}
The CNN is a feedforward deep learning model. In the convolutional layer, the fused feature matrix of input texture features and opcode features is divided into multiple submatrices for convolution operations. As shown in (4)
, \( Z \) represents the result after convolution, \( W \) denotes the weights, and \( b \) is the bias.
\begin{equation}
    Z=f(MW+b)\label{eq}
\end{equation}
As shown in (5), the activation function Relu is used for nonlinear mapping.
\begin{equation}
    f=relu=max(0,x)\label{eq}
\end{equation}
After the previous step, the processed results are passed to the pooling layer for filtering. By reducing the number of model parameters, this step decreases computational complexity and also helps prevent overfitting. In the CNN layer, through convolution and pooling operations, the deep features of malicious code are gradually enhanced and preserved. The strengthened features are then fed into the BiLSTM model for the next phase of learning and training.
\subsubsection{BiLSTM Layer}
After passing through the previous CNN layer, the enhanced features of the malicious code are obtained. These enhanced features serve as input to the BiLSTM model. The input gate is defined by (6).
\begin{equation}
    i_{t}  =\sigma \left ( W_{xi}x_{t}+W_{hi} h_{t-1} +W_{ci} c_{t-1} +b_{i}    \right ) \label{eq}
\end{equation}
The forget gate receives the output from the CNN layer and determines whether to forget the information.
\begin{equation}
f_{t} =\sigma \left ( W_{xf}x_{t} +W_{hf}h_{t-1}   +W_{cf}c_{t-1}  +b_{f}  \right ) \label{eq}
\end{equation}
At this point, the cell state receives inputs from both the input gate and the forget gate.
\begin{equation}
    c_{t}=f_{c}\odot c_{t-1}   +i_{t} \odot tanh\left ( W_{xc}x_{t} +W_{hc}h_{t-1}  +b_{c}   \right ) \label{eq}
\end{equation}
The output gate controls the output of the cell state.
\begin{equation}
    o_{t} =\sigma \left ( W_{xo}x_{t}+W_{ho}h_{t-1}+W_{co}c_{t-1} +b_{o}       \right ) \label{eq}
\end{equation}
The output of the current cell is given by (9).
\begin{equation}
    h_{t}=o_{t}\odot \tan c_{t}   \label{eq}
\end{equation}
Here, \( f_t \) represents the forget gate, \( i_t \) denotes the input gate, and \( o_t \) stands for the output gate. \( W_f , W_i , W_C , W_o , b_f , b_i , b_c , \) and \( b_o \) are the weights and biases. \( c_t \) is the cell state, and \( h_t \) is the hidden output.
For the sentence \( x = [x_1, x_2, \ldots, x_n] \), the forward LSTM generates a hidden state sequence \( \overrightarrow{h_s} \in \mathbb{R}^{n \times d_h} \) from the feature vector output by the CNN, while the backward LSTM also generates a hidden state sequence \(\overleftarrow{h_s} \in \mathbb{R}^{n \times d_h} \) from the same CNN output. The final output hidden state sequence \( \overrightarrow{h_s} \in \mathbb{R}^{n \times 2d_h} \) is composed of the sequences from both the forward and backward outputs.
\begin{equation}
\overrightarrow{h_s} = \text{LSTM}([v_1; v_2; \ldots; v_n])
\label{eq}
\end{equation}
\begin{equation}
    \overleftarrow{h_s} = \text{LSTM}([v_1; v_2; \ldots; v_n])
\label{eq}
\end{equation}
\begin{equation}
h_s = [\overrightarrow{h_s}, \overleftarrow{h_s})\label{eq}
\end{equation}
The Adam algorithm is characterized by its ease of implementation, efficiency, and low memory usage. Therefore, the Adam optimizer is used to optimize the gradient descent. Based on the fused features of texture features and opcode features, along with the CNN-BiLSTM model, the process is illustrated in Algorithm 3.
\begin{algorithm}
\caption{CNN-BiLSTM Model}
\KwIn{$X_n^{API}$ (fused features), $Y_n^{API}$ (labels)}
\KwOut{Fused Model (CNN-BiLSTM)}

1. Define neural network layers: $CNN\_layers = layer1$, $Bi-LSTM\_layers = layer2$ \;
2. Set the number of folds $K = 10$ \;
3. Set the number of training iterations $N = 10$ \;
4. \For{$k \in \text{range}(1, K, 1)$}{
    5. \For{$n \in \text{range}(1, N, 1)$}{
        6. \If{$this.layer[layer1] == CNN$}{
            7. random($W, b$) // randomly initialize \;
            8. Perform convolution operation \;
        }
        9. \If{$this.layer[layer1] == MaxPooling$}{
            10. Perform pooling operation \;
        }
        11. \Else{
            12. Enter fully connected layer, reducing dimensions to one \;
        }
        13. End if \;
        14. \If{$this.layer[layer2] == Bi-LSTM$}{
            15. random($w, b$) // randomly initialize \;
            16. Compute hidden layer of Bi-LSTM \;
            17. Output the fused model (CNN-BiLSTM) \;
        }
        18. End if \;
    }
}
\end{algorithm}

The CNN layers learn and enhance the local structural features of the fused features, which are then input into the BiLSTM to further explore the contextual relationship information of the fused features. Therefore, utilizing the CNN-BiLSTM fusion model allows for improved detection and classification of malware. The structure of the CNN-BiLSTM fusion model is illustrated in Figure 6.
\begin{figure}[H]
    \centering
    \includegraphics[width=0.9\linewidth]{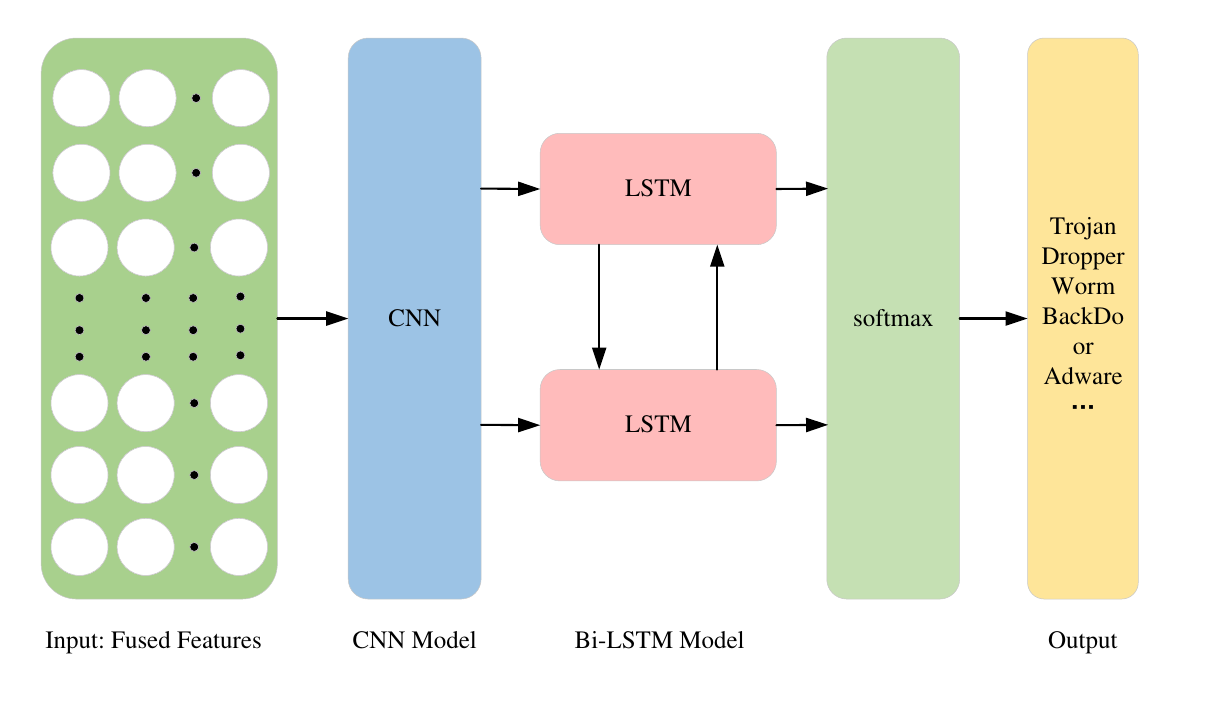}
    \caption{CNN-BiLSTM Model Structure Diagram}
    \label{fig:enter-label}
\end{figure}

\section{Experiments and Results}
The previous section provided a detailed explanation of how to conduct static malware detection. Initially, the MinHash algorithm is utilized to generate grayscale images, followed by the extraction of global and local features using the GIST and LBP algorithms to create a hybrid feature set. Subsequently, opcode features are extracted using the N-gram and tf-idf algorithms. The texture features and opcode features are then fused together, and the combined features are input into the CNN-BiLSTM model for training.

By following these steps, a static malware detection model can be successfully constructed. However, to enhance the model's adaptability to the fused features, parameter tuning is necessary. Regarding the value of n in the N-gram algorithm, it is set within the range of 1 to 5, and the optimal n value is determined through experimental testing. The experimental results, as shown in Table 2, indicate that the highest accuracy is achieved when n=4.
\begin{table}[H]
\centering
\caption{Comparison of Experimental Results with Different N Values}
\label{tab:my-table}
\begin{tabular}{|c|c|c|}
\hline
\textbf{Serial Number} & \textbf{N-gram with Different N Values} & \textbf{Accuracy (ACC)} \\ \hline
1                      & 1-gram+CNN-BiLSTM                       & 0.850                   \\ \hline
2                      & 2-gram+CNN-BiLSTM                       & 0.874                   \\ \hline
3                      & 3-gram+CNN-BiLSTM                       & 0.917                   \\ \hline
4                      & 4-gram+CNN-BiLSTM                       & 0.940                   \\ \hline
5                      & 5-gram+CNN-BiLSTM                       & 0.923                   \\ \hline
\end{tabular}
\end{table}
In this paper, the GIST algorithm was utilized to obtain global texture features from grayscale images, while the LBP algorithm was employed to extract local texture features. The two types of texture features were then combined and input into the CNN-BiLSTM model for training. Experiments were conducted using both the individual texture features and the combined texture features, with the results presented in Table 3. It can be observed from the table that the accuracy achieved by combining the two texture features is higher than that obtained by training with a single texture feature.
\begin{table}[H]
\centering
\caption{Results for Different Texture Feature Selections}
\resizebox{\linewidth}{!}{
\label{tab:my-table}
\begin{tabular}{|c|c|c|}
\hline
\textbf{Serial Number} & \textbf{Method for Inputting Different Texture Features} & \textbf{Accuracy (ACC)} \\ \hline
1                      & GIST+CNN-BiLSTM                                          & 0.929                   \\ \hline
2                      & LBP+CNN-BiLSTM                                           & 0.950                   \\ \hline
3                      & Combined Texture Features+CNN-BiLSTM                     & 0.966                   \\ \hline
\end{tabular}
}
\end{table}
A comparison is made between the use of individual opcode features or texture features and the use of fused features, with the model employing the hybrid CNN-BiLSTM architecture. The experimental process is illustrated in Figure 7.
\begin{figure}[H]
    \centering
    \includegraphics[width=0.7\linewidth]{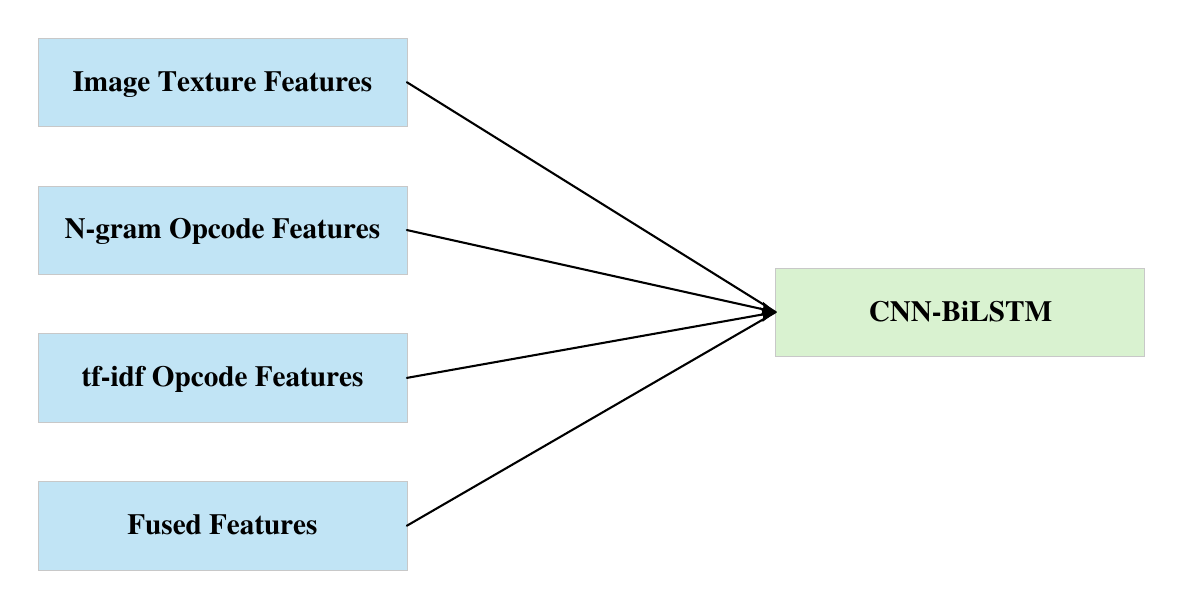}
    \caption{Experimental Process for Comparing the Advantages and Disadvantages of Features}
    \label{fig:enter-label}
\end{figure}
The experimental results, presented in Table 4, show that the classification accuracy for opcode features extracted using the N-gram algorithm is 93.9\%, while tf-idf achieves 94.3\%. Texture features have an accuracy of 96.6\%, and fused features reach 98.7\%. The results indicate that the model utilizing fused features outperforms those using individual feature types. Fused features capture malware's behavioral characteristics from multiple dimensions, enhancing detection accuracy, whereas single-level features only describe behavior from local or global scales, lacking comprehensive representation.
\begin{table}[H]
\centering
\caption{Results Using Different Features}
\resizebox{\linewidth}{!}{
\label{tab:my-table}
\begin{tabular}{|c|c|c|}
\hline
\textbf{Serial Number} & \textbf{Selected Different Features as Input} & \textbf{Accuracy (ACC)} \\ \hline
1                      & N-gram Opcode Features + CNN-BiLSTM           & 0.939                   \\ \hline
2                      & tf-idf Opcode Features + CNN-BiLSTM           & 0.943                   \\ \hline
3                      & Image Texture Features + CNN-BiLSTM           & 0.966                   \\ \hline
4                      & Fused Features + CNN-BiLSTM                   & 0.987                   \\ \hline
\end{tabular}
}
\end{table}
A comparative experiment is conducted among traditional models, deep learning models, and the fusion model presented in this paper. Accuracy, precision, recall, and F1-Score are used as evaluation metrics.. Six models are compared: Support Vector Machine (SVM), Logistic Regression, Naive Bayes, Convolutional Neural Network (CNN), Long Short-Term Memory (LSTM), and the CNN-BiLSTM fusion model. All models use fused features as input, and the N-gram algorithm consistently employs the 4-gram variant.The experimental results are presented in Table 5.
\begin{table}[H]
\centering
\caption{Comparison of Experimental Results for Six Models}
\resizebox{\linewidth}{!}{
\label{tab:my-table}
\begin{tabular}{|c|c|c|c|c|c|}
\hline
\textbf{Serial Number} & \textbf{Model Name}                & \textbf{Precision} & Recall & F1 Score & Accuracy (ACC) \\ \hline
1                      & Support Vector Machine (SVM)       & 0.933              & 0.892  & 0.915    & 0.930          \\ \hline
2                      & Logistic Regression (LR)           & 0.868              & 0.841  & 0.856    & 0.916          \\ \hline
3                      & Naive Bayes (NB)                   & 0.955              & 0.848  & 0.869    & 0.899          \\ \hline
4                      & Convolutional Neural Network (CNN) & 0.932              & 0.909  & 0.912    & 0.933          \\ \hline
5                      & Long Short-Term Memory (LSTM)      & 0.952              & 0.937  & 0.921    & 0.962          \\ \hline
6                      & Fusion Model (CNN-BiLSTM)          & 0.961              & 0.944  & 0.943    & 0.987          \\ \hline
\end{tabular}
}
\end{table}
The accuracy of these six models trained ten times is shown in Figure 8.
\begin{figure}[H]
    \centering
    \includegraphics[width=0.8\linewidth]{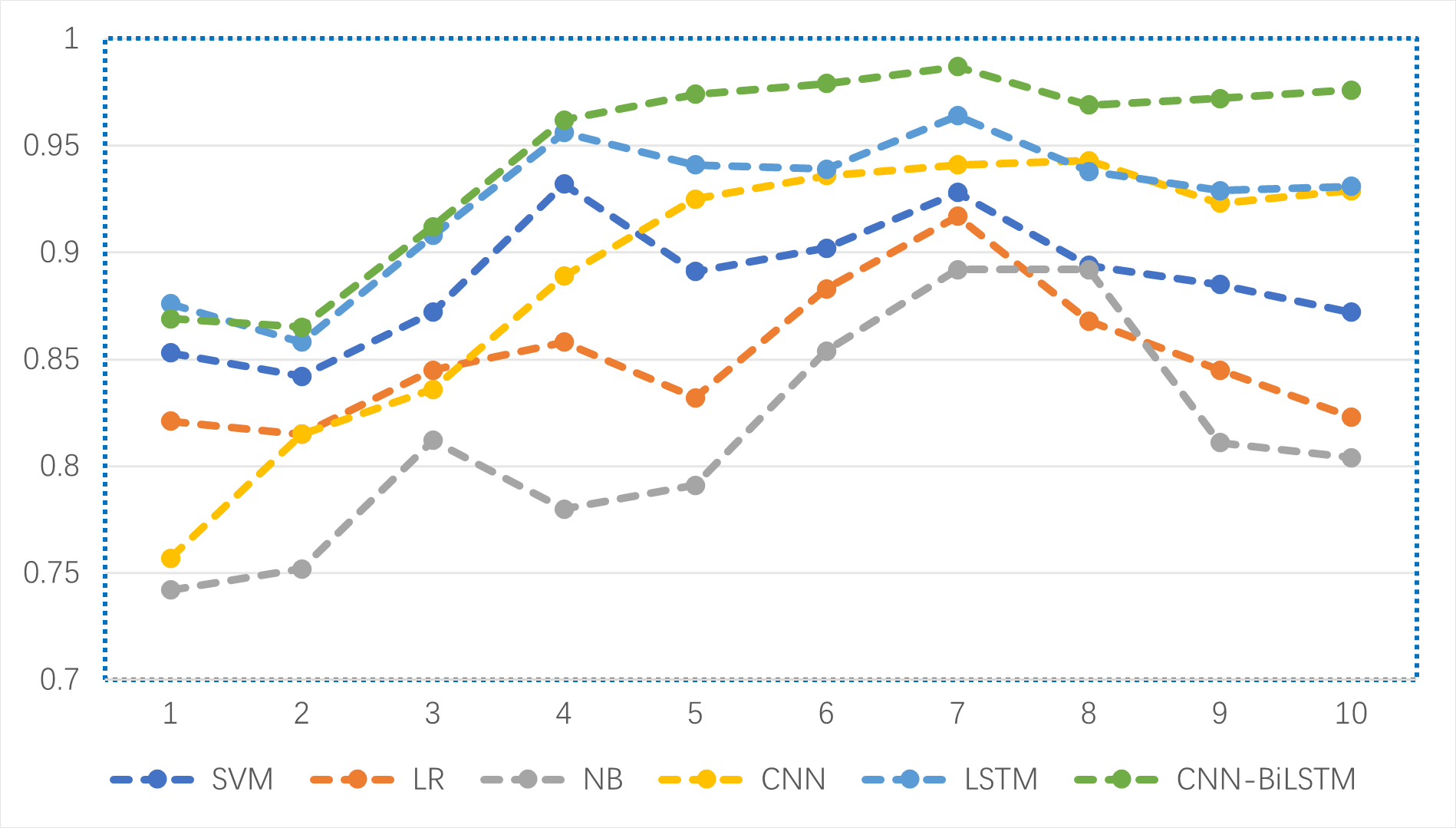}
    \caption{Line Chart of Training Accuracy}
    \label{fig:enter-label}
\end{figure}

From the graphical and tabular experimental results, it can be observed that the fusion model CNN-BiLSTM outperforms or equals other models in all metrics. Its accuracy is higher by 0.6\% to 9.3\% compared to other models, recall is higher by 0.7\% to 10.3\%, F1 score is higher by 2.2\% to 8.7\%, and accuracy is higher by 1.5\% to 8.8\%. This indicates that the classification performance of the CNN-BiLSTM model using fused features is superior to that of other models.

When comparing the method used in this paper with the methods from referenced literature, the results are presented in Table 6. It can be seen that the method employed in this study achieves higher accuracy than the ResNet and GoogleNet models from ref.\cite{liu2025pmanet}, higher than the CNN model from ref.\cite{kim2018image}, and higher than the malicious code visualization method from ref.\cite{yadav2023malware}. This comparison demonstrates that the method based on fused features and the fusion model CNN-BiLSTM used in this paper exhibits superior performance.
\begin{table}[H]
\centering
\caption{Comparison of Results with Existing Approaches}
\label{tab:my-table}
\begin{tabular}{|c|c|c|}
\hline
\textbf{Serial Number} & \textbf{Recognition Method Name} & \textbf{Accuracy (ACC)} \\ \hline
1                      & ResNet Model{[}59{]}             & 0.884                   \\ \hline
2                      & GoogleNet Model{[}59{]}          & 0.745                   \\ \hline
3                      & CNN-image Model{[}60{]}          & 0.917                   \\ \hline
4                      & Visualization Method{[}61{]}     & 0.983                   \\ \hline
5                      & Fusion Features + CNN-BiLSTM     & 0.987                   \\ \hline
\end{tabular}
\end{table}
The experimental results demonstrate that feature fusion and the CNN-BiLSTM model significantly enhance malicious code detection. By integrating texture features (GIST, LBP) with opcode features (N-gram, tf-idf), classification accuracy improves by 2.1\%-4.6\% over single features. The CNN-BiLSTM model further boosts accuracy to 98.7\%, surpassing traditional models (SVM, LSTM) by 2.3\%-5.2\%. Compared to other models (e.g., ResNet, GoogleNet), this approach excels in accuracy and variant detection, highlighting the effectiveness of combining feature fusion with deep learning. The fusion of features provides stronger abstraction, while the CNN-BiLSTM model captures deeper contextual information, resulting in superior detection performance.

\section{Conclusion}
We present a novel approach to malicious code detection using feature fusion and a CNN-BiLSTM deep learning model. By integrating texture features (GIST and LBP) with opcode features (N-gram and tf-idf), we address the limitations of single-feature models, improving classification accuracy by 2.1\%-4.6\%. The CNN-BiLSTM model captures both spatial and sequential relationships, achieving an accuracy of 98.7\%, which is 2.3\%-5.2\% higher than traditional models like SVM and LSTM, and outperforms state-of-the-art models such as ResNet and GoogleNet in detecting malicious code variants.

Future research should focus on enhancing model efficiency and reducing training time through techniques like transfer learning and model pruning. Incorporating diverse feature sets, including behavioral data, will strengthen detection against sophisticated threats. Adapting the model for various platforms, especially mobile and IoT devices, is essential as attackers increasingly target these environments. Additionally, exploring explainable AI (XAI) will enhance the interpretability of detection results, fostering trust in automated systems.

In summary, combining feature fusion with deep learning provides an effective strategy for advancing malicious code detection, paving the way for future innovations to combat evolving cyber threats.

\vspace{12pt}


\begin{thebibliography}{10}

\bibitem{lis2019cyberattacks}
P.~Lis and J.~Mendel, ``Cyberattacks on critical infrastructure: An economic perspective,'' {\em Economics and Business Review}, vol.~5, no.~2, pp.~24--47, 2019.

\bibitem{kumar2023emerging}
I.~Kumar {\em et~al.}, ``Emerging threats in cybersecurity: a review article,'' {\em International Journal of Applied and Natural Sciences}, vol.~1, no.~1, pp.~01--08, 2023.

\bibitem{lehto2022cyber}
M.~Lehto, ``Cyber-attacks against critical infrastructure,'' in {\em Cyber security: Critical infrastructure protection}, pp.~3--42, Springer, 2022.

\bibitem{thakur2024cyber}
M.~Thakur, ``Cyber security threats and countermeasures in digital age,'' {\em Journal of Applied Science and Education (JASE)}, vol.~4, no.~1, pp.~1--20, 2024.

\bibitem{sihwail2018survey}
R.~Sihwail, K.~Omar, and K.~Z. Ariffin, ``A survey on malware analysis techniques: Static, dynamic, hybrid and memory analysis,'' {\em Int. J. Adv. Sci. Eng. Inf. Technol}, vol.~8, no.~4-2, pp.~1662--1671, 2018.

\bibitem{10.1007/978-3-642-20267-4_11}
H.~Parvin, B.~Minaei, H.~Karshenas, and A.~Beigi, ``A new n-gram feature extraction-selection method for malicious code,'' in {\em Adaptive and Natural Computing Algorithms} (A.~Dobnikar, U.~Lotri{\v{c}}, and B.~{\v{S}}ter, eds.), (Berlin, Heidelberg), pp.~98--107, Springer Berlin Heidelberg, 2011.

\bibitem{malware2011obfuscation}
B.~Malware, ``Obfuscation: The hidden malware,'' {\em IEEE Security \& Privacy}, 2011.

\bibitem{wolsey2022stateoftheartaibasedmalwaredetection}
A.~Wolsey, ``The state-of-the-art in ai-based malware detection techniques: A review,'' 2022.

\bibitem{app12178482}
F.~A. Aboaoja, A.~Zainal, F.~A. Ghaleb, B.~A.~S. Al-rimy, T.~A.~E. Eisa, and A.~A.~H. Elnour, ``Malware detection issues, challenges, and future directions: A survey,'' {\em Applied Sciences}, vol.~12, no.~17, 2022.

\bibitem{li2020advanced}
Z.~Li, {\em Advanced Techniques to Detect Complex Android Malware}.
\newblock PhD thesis, The University of Nebraska-Lincoln, 2020.

\bibitem{bacci2018impact}
A.~Bacci, A.~Bartoli, F.~Martinelli, E.~Medvet, F.~Mercaldo, C.~A. Visaggio, {\em et~al.}, ``Impact of code obfuscation on android malware detection based on static and dynamic analysis.,'' in {\em ICISSP}, pp.~379--385, 2018.

\bibitem{dalla2007code}
M.~Dalla~Preda {\em et~al.}, ``Code obfuscation and malware detection by abstract interpretation,'' 2007.

\bibitem{Lin2018Efficient}
C.~Lin, H.~Pao, and J.-W. Liao, ``Efficient dynamic malware analysis using virtual time control mechanics,'' {\em Comput. Secur.}, vol.~73, pp.~359--373, 2018.

\bibitem{Ngo2022Fast}
M.~V. Ngo, T.~Truong-Huu, D.~Rabadi, J.~Loo, and S.~Teo, ``Fast and efficient malware detection with joint static and dynamic features through transfer learning,'' pp.~503--531, 2022.

\bibitem{Carlin2019A}
D.~Carlin, P.~O'Kane, and S.~Sezer, ``A cost analysis of machine learning using dynamic runtime opcodes for malware detection,'' {\em Comput. Secur.}, vol.~85, pp.~138--155, 2019.

\bibitem{Mills2020Investigating}
A.~Mills and P.~Legg, ``Investigating anti-evasion malware triggers using automated sandbox reconfiguration techniques,'' {\em Journal of Cybersecurity and Privacy}, 2020.

\bibitem{Aboaoja2023Dynamic}
F.~A. Aboaoja, A.~Zainal, A.~Ali, F.~A. Ghaleb, F.~Alsolami, and M.~Rassam, ``Dynamic extraction of initial behavior for evasive malware detection,'' {\em Mathematics}, 2023.

\bibitem{Brown2023Automated}
A.~R. Brown, M.~Gupta, and M.~Abdelsalam, ``Automated machine learning for deep learning based malware detection,'' {\em ArXiv}, vol.~abs/2303.01679, 2023.

\bibitem{Vinayakumar2019Robust}
R.~Vinayakumar, M.~Alazab, K.~Soman, P.~Poornachandran, and S.~Venkatraman, ``Robust intelligent malware detection using deep learning,'' {\em IEEE Access}, vol.~7, pp.~46717--46738, 2019.

\bibitem{Shukla2019Stealthy}
S.~Shukla, G.~Kolhe, S.~M.~P. Dinakarrao, and S.~Rafatirad, ``Stealthy malware detection using rnn-based automated localized feature extraction and classifier,'' {\em 2019 IEEE 31st International Conference on Tools with Artificial Intelligence (ICTAI)}, pp.~590--597, 2019.

\bibitem{Kim2022FILM:}
Y.~J. Kim, C.-H. Park, and M.~Yoon, ``Film: Filtering and machine learning for malware detection in edge computing,'' {\em Sensors (Basel, Switzerland)}, vol.~22, 2022.

\bibitem{Al-Hashmi2022Deep-Ensemble}
A.~A. Al-Hashmi, F.~A. Ghaleb, A.~Al-Marghilani, A.~E. Yahya, S.~A. Ebad, M.~S. M.S., and A.~A. Darem, ``Deep-ensemble and multifaceted behavioral malware variant detection model,'' {\em IEEE Access}, vol.~10, pp.~42762--42777, 2022.

\bibitem{Hemalatha2021An}
J.~Hemalatha, S.~A. Roseline, S.~Geetha, S.~Kadry, and R.~Damaševičius, ``An efficient densenet-based deep learning model for malware detection,'' {\em Entropy}, vol.~23, 2021.

\bibitem{Gibert2022Fusing}
D.~Gibert, C.~Mateu, J.~Planes, and Q.~Le, ``Fusing feature engineering and deep learning: A case study for malware classification,'' {\em Expert Syst. Appl.}, vol.~207, p.~117957, 2022.

\bibitem{Yang2023A}
X.~Yang, D.~Yang, and Y.~Li, ``A hybrid attention network for malware detection based on multi-feature aligned and fusion,'' {\em Electronics}, 2023.

\bibitem{Ding2019A}
Y.~Ding, J.~Hu, W.~Xu, and X.~Zhang, ``A deep feature fusion method for android malware detection,'' {\em 2019 International Conference on Machine Learning and Cybernetics (ICMLC)}, pp.~1--6, 2019.

\bibitem{Kim2019A}
T.~Kim, B.~Kang, M.~Rho, S.~Sezer, and E.~Im, ``A multimodal deep learning method for android malware detection using various features,'' {\em IEEE Transactions on Information Forensics and Security}, vol.~14, pp.~773--788, 2019.

\bibitem{jalilian2020static}
A.~Jalilian, Z.~Narimani, and E.~Ansari, ``Static signature-based malware detection using opcode and binary information,'' in {\em Data Science: From Research to Application}, pp.~24--35, Springer, 2020.

\bibitem{10.1007/978-3-030-37309-2_3}
A.~Jalilian, Z.~Narimani, and E.~Ansari, ``Static signature-based malware detection using opcode and binary information,'' in {\em Data Science: From Research to Application} (M.~Bohlouli, B.~Sadeghi~Bigham, Z.~Narimani, M.~Vasighi, and E.~Ansari, eds.), (Cham), pp.~24--35, Springer International Publishing, 2020.

\bibitem{info15020102}
E.~Alkhateeb, A.~Ghorbani, and A.~Habibi~Lashkari, ``Identifying malware packers through multilayer feature engineering in static analysis,'' {\em Information}, vol.~15, no.~2, 2024.

\bibitem{sun2019opcode}
Z.~Sun, Z.~Rao, J.~Chen, R.~Xu, D.~He, H.~Yang, and J.~Liu, ``An opcode sequences analysis method for unknown malware detection,'' in {\em Proceedings of the 2019 2nd international conference on geoinformatics and data analysis}, pp.~15--19, 2019.

\bibitem{santos2013opcode}
I.~Santos, F.~Brezo, X.~Ugarte-Pedrero, and P.~G. Bringas, ``Opcode sequences as representation of executables for data-mining-based unknown malware detection,'' {\em information Sciences}, vol.~231, pp.~64--82, 2013.

\bibitem{faruki2023survey}
P.~Faruki, R.~Bhan, V.~Jain, S.~Bhatia, N.~El~Madhoun, and R.~Pamula, ``A survey and evaluation of android-based malware evasion techniques and detection frameworks,'' {\em Information}, vol.~14, no.~7, p.~374, 2023.

\bibitem{jamalpur2018dynamic}
S.~Jamalpur, Y.~S. Navya, P.~Raja, G.~Tagore, and G.~R.~K. Rao, ``Dynamic malware analysis using cuckoo sandbox,'' in {\em 2018 Second international conference on inventive communication and computational technologies (ICICCT)}, pp.~1056--1060, IEEE, 2018.

\bibitem{or2019dynamic}
O.~Or-Meir, N.~Nissim, Y.~Elovici, and L.~Rokach, ``Dynamic malware analysis in the modern era—a state of the art survey,'' {\em ACM Computing Surveys (CSUR)}, vol.~52, no.~5, pp.~1--48, 2019.

\bibitem{ilic2022pilot}
S.~{\v{Z}}. Ili{\'c}, M.~J. Gnjatovi{\'c}, B.~M. Popovi{\'c}, and N.~D. Ma{\v{c}}ek, ``A pilot comparative analysis of the cuckoo and drakvuf sandboxes: An end-user perspective,'' {\em Vojnotehni{\v{c}}ki glasnik/Military Technical Courier}, vol.~70, no.~2, pp.~372--392, 2022.

\bibitem{maniriho2022study}
P.~Maniriho, A.~N. Mahmood, and M.~J.~M. Chowdhury, ``A study on malicious software behaviour analysis and detection techniques: Taxonomy, current trends and challenges,'' {\em Future Generation Computer Systems}, vol.~130, pp.~1--18, 2022.

\bibitem{singh2021survey}
J.~Singh and J.~Singh, ``A survey on machine learning-based malware detection in executable files,'' {\em Journal of Systems Architecture}, vol.~112, p.~101861, 2021.

\bibitem{gopinath2023comprehensive}
M.~Gopinath and S.~C. Sethuraman, ``A comprehensive survey on deep learning based malware detection techniques,'' {\em Computer Science Review}, vol.~47, p.~100529, 2023.

\bibitem{9110843}
K.~Saeed, W.~Khalil, S.~Ahmed, I.~Ahmad, and M.~N.~K. Khattak, ``Seecr: Secure energy efficient and cooperative routing protocol for underwater wireless sensor networks,'' {\em IEEE Access}, vol.~8, pp.~107419--107433, 2020.

\bibitem{hashemi2022ensemble}
A.~Hashemi, M.~B. Dowlatshahi, and H.~Nezamabadi-pour, ``Ensemble of feature selection algorithms: a multi-criteria decision-making approach,'' {\em International Journal of Machine Learning and Cybernetics}, vol.~13, no.~1, pp.~49--69, 2022.

\bibitem{shiri2023comprehensiveoverviewcomparativeanalysis}
F.~M. Shiri, T.~Perumal, N.~Mustapha, and R.~Mohamed, ``A comprehensive overview and comparative analysis on deep learning models: Cnn, rnn, lstm, gru,'' 2023.

\bibitem{Omar2022}
M.~Omar, {\em New Approach to Malware Detection Using Optimized Convolutional Neural Network}, pp.~13--35.
\newblock Cham: Springer International Publishing, 2022.

\bibitem{app13074624}
B.~Saridou, I.~Moulas, S.~Shiaeles, and B.~Papadopoulos, ``Image-based malware detection using α-cuts and binary visualisation,'' {\em Applied Sciences}, vol.~13, no.~7, 2023.

\bibitem{gibert2019using}
D.~Gibert, C.~Mateu, J.~Planes, and R.~Vicens, ``Using convolutional neural networks for classification of malware represented as images,'' {\em Journal of Computer Virology and Hacking Techniques}, vol.~15, pp.~15--28, 2019.

\bibitem{10.1007/978-3-030-16657-1_9}
J.~Mathew and M.~A. Ajay~Kumara, ``Api call based malware detection approach using recurrent neural network---lstm,'' in {\em Intelligent Systems Design and Applications} (A.~Abraham, A.~K. Cherukuri, P.~Melin, and N.~Gandhi, eds.), (Cham), pp.~87--99, Springer International Publishing, 2020.

\bibitem{xiao2019android}
X.~Xiao, S.~Zhang, F.~Mercaldo, G.~Hu, and A.~K. Sangaiah, ``Android malware detection based on system call sequences and lstm,'' {\em Multimedia Tools and Applications}, vol.~78, pp.~3979--3999, 2019.

\bibitem{electronics13132539}
H.~Kim and M.~Kim, ``Malware detection and classification system based on cnn-bilstm,'' {\em Electronics}, vol.~13, no.~13, 2024.

\bibitem{johny2024deep}
J.~A. Johny, G.~Radhamani, M.~Conti, {\em et~al.}, ``Deep learning fusion for effective malware detection: Leveraging visual features,'' {\em arXiv preprint arXiv:2405.14311}, 2024.

\bibitem{math11132944}
A.~A. Alhashmi, A.~A. Darem, A.~M. Alashjaee, S.~M. Alanazi, T.~M. Alkhaldi, S.~A. Ebad, F.~A. Ghaleb, and A.~M. Almadani, ``Similarity-based hybrid malware detection model using api calls,'' {\em Mathematics}, vol.~11, no.~13, 2023.

\bibitem{10.1007/978-981-19-7346-8_63}
O.~P. Samantray and S.~N. Tripathy, ``An efficient hybrid approach for malware detection using frequent opcodes and api call sequences,'' in {\em Computational Intelligence} (A.~Shukla, B.~K. Murthy, N.~Hasteer, and J.-P. Van~Belle, eds.), (Singapore), pp.~727--735, Springer Nature Singapore, 2023.

\bibitem{10.1007/978-3-030-91356-4_14}
M.~R. Norouzian, P.~Xu, C.~Eckert, and A.~Zarras, ``Hybroid: Toward android malware detection and categorization with program code and network traffic,'' in {\em Information Security} (J.~K. Liu, S.~Katsikas, W.~Meng, W.~Susilo, and R.~Intan, eds.), (Cham), pp.~259--278, Springer International Publishing, 2021.

\bibitem{damodaran2017comparison}
A.~Damodaran, F.~D. Troia, C.~A. Visaggio, T.~H. Austin, and M.~Stamp, ``A comparison of static, dynamic, and hybrid analysis for malware detection,'' {\em Journal of Computer Virology and Hacking Techniques}, vol.~13, pp.~1--12, 2017.

\bibitem{zhu2021malware}
X.~Zhu, J.~Huang, B.~Wang, and C.~Qi, ``Malware homology determination using visualized images and feature fusion,'' {\em PeerJ Computer Science}, vol.~7, p.~e494, 2021.

\bibitem{hashemi2023ifmd}
H.~Hashemi, M.~E. Samie, and A.~Hamzeh, ``Ifmd: image fusion for malware detection,'' {\em Journal of Computer Virology and Hacking Techniques}, vol.~19, no.~2, pp.~271--286, 2023.

\bibitem{10.1007/978-3-031-49099-6_10}
N.~Singh and S.~Tripathy, ``Mdldroid: Multimodal deep learning based android malware detection,'' in {\em Information Systems Security} (V.~Muthukkumarasamy, S.~D. Sudarsan, and R.~K. Shyamasundar, eds.), (Cham), pp.~159--177, Springer Nature Switzerland, 2023.

\bibitem{akhtar2022detection}
M.~S. Akhtar and T.~Feng, ``Detection of malware by deep learning as cnn-lstm machine learning techniques in real time,'' {\em Symmetry}, vol.~14, no.~11, p.~2308, 2022.

\bibitem{jeon2021hybrid}
J.~Jeon, B.~Jeong, S.~Baek, and Y.-S. Jeong, ``Hybrid malware detection based on bi-lstm and spp-net for smart iot,'' {\em IEEE Transactions on Industrial Informatics}, vol.~18, no.~7, pp.~4830--4837, 2021.

\bibitem{khan2019analysis}
R.~U. Khan, X.~Zhang, and R.~Kumar, ``Analysis of resnet and googlenet models for malware detection,'' {\em Journal of Computer Virology and Hacking Techniques}, vol.~15, pp.~29--37, 2019.

\bibitem{kim2018image}
H.-J. Kim, ``Image-based malware classification using convolutional neural network,'' in {\em Advances in Computer Science and Ubiquitous Computing: CSA-CUTE 17}, pp.~1352--1357, Springer, 2018.

\bibitem{yadav2023malware}
B.~Yadav and S.~Tokekar, ``Malware multi-class classification based on malware visualization using a convolutional neural network model,'' {\em International Journal of Information Engineering and Electronic Business}, vol.~15, no.~2, p.~20, 2023.

\bibitem{liu2025pmanet}
R.~Liu, Y.~Wang, H.~Xu, Z.~Qin, F.~Zhang, Y.~Liu, and Z.~Cao, ``Pmanet: Malicious url detection via post-trained language model guided multi-level feature attention network,'' {\em Information Fusion}, vol.~113, p.~102638, 2025.

\bibitem{liu2024transurl}
R.~Liu, Y.~Wang, Z.~Guo, H.~Xu, Z.~Qin, W.~Ma, and F.~Zhang, ``Transurl: Improving malicious url detection with multi-layer transformer encoding and multi-scale pyramid features,'' {\em Computer Networks}, vol.~253, p.~110707, 2024.

\bibitem{qian2024mufuzz}
P.~Qian, H.~Wu, Z.~Du, T.~Vural, D.~Rong, Z.~Cao, L.~Zhang, Y.~Wang, J.~Chen, and Q.~He, ``Mufuzz: Sequence-aware mutation and seed mask guidance for blockchain smart contract fuzzing,'' in {\em 2024 IEEE 40th International Conference on Data Engineering (ICDE)}, pp.~1972--1985, IEEE, 2024.

\bibitem{pu2024larod}
Y.~Pu, C.~Gao, B.~Li, S.~Liu, S.~Yang, J.~Xiao, S.~Pu, Z.~You, and Y.~Wang, ``Larod-hd: Low-cost adaptive real-time object detection for high-resolution video surveillance,'' in {\em International Conference on Intelligent Computing}, pp.~253--265, Springer, 2024.

\bibitem{jia2024ethereum}
Y.~Jia, Y.~Wang, J.~Sun, Y.~Liu, Z.~Sheng, and Y.~Tian, ``Ethereum fraud detection via joint transaction language model and graph representation learning,'' {\em arXiv preprint arXiv:2409.07494}, 2024.

\end{thebibliography}
\end{document}